\documentclass[a4paper,11pt]{article}
\usepackage{geometry}
\usepackage[T1]{fontenc}
\usepackage[utf8]{inputenc}
\usepackage{lmodern}
\usepackage{graphicx}
\usepackage{color}
\usepackage{amsmath,amsfonts,amssymb}
\usepackage[colorlinks=true, urlcolor=blue]{hyperref}
\everymath{\displaystyle}
\usepackage{authblk}

\title{\textbf{$\mathbf{U(1)_A}$ axial anomaly, $\mathbf{\eta^\prime}$, and topological susceptibility in the holographic soft-wall model}}
\author[1]{Floriana Giannuzzi\thanks{\href{{mailto:floriana.giannuzzi@ba.infn.it}}{floriana.giannuzzi@ba.infn.it}}}
\author[1]{Stefano Nicotri\thanks{\href{mailto:nicotri@infn.it}{nicotri@infn.it}}}
\affil[1]{\small \emph{INFN -- Istituto Nazionale di Fisica Nucleare -- Sezione di Bari} \protect\\ \emph{Via Orabona 4, 70125, Bari, Italy}}
\date{}

\begin{document}
\begin{flushright}{BARI-TH/21-727} \end{flushright}
{\let\newpage\relax\maketitle}
\maketitle

\begin{abstract}
    The singlet pseudoscalar sector of light mesons is investigated in the soft-wall model, a bottom-up approach to the AdS/QCD correspondence. The $\eta^\prime$ mass results from the mixing among the fields dual to the axial current, pseudoscalar current and the $G\tilde G$ operator. The topological susceptibility is computed for any quark mass, and the Witten-Veneziano relation is obtained in the large $N_c$ limit.
\end{abstract}

\section{Introduction}
In QCD the absence of a singlet pseudo-Goldostone boson is the well-known $U(1)_A$ problem. 
If the $U(1)_A$ symmetry were spontaneously broken, a very light isosinglet pseudoscalar Goldstone boson would be generated, with a mass $\sim \sqrt{3} m_\pi$ according to chiral perturbation theory \cite{Weinberg:1975ui}. 
However, although there is no conserved $U(1)_A$ quantum number, an extra Goldstone boson is missing: the mass of the $\eta^\prime$ meson, the candidate for such a state, is 958 MeV \cite{Zyla:2020zbs}, significantly higher than the predicted value.
$U(1)_A$ is anomalous, and the nonconservation of the singlet axial current in the chiral limit is expressed by the anomaly equation:
\begin{equation}
\partial_\mu J_A^\mu = - \frac{\alpha_s}{8 \pi} \, G_{\mu\nu} \tilde G^{\mu\nu}
\end{equation}
where $G_{\mu\nu}$ is the gluon field strength and $\tilde G^{\mu\nu} = \epsilon^{\mu\nu\alpha\beta} G_{\alpha\beta}/2$.
In the past there have been some discussions on the way the Goldstone boson receives a mass as a result of the anomaly.
't Hooft proposed that the violation of $U(1)_A$ is realized by instanton configurations that explicitly break the symmetry and  contribute to the $\eta^\prime$ mass \cite{tHooft:1976rip}\cite{tHooft:1986ooh}. 
Different hypotheses were also put forward. 
In some quark models the high mass of the singlet state is caused by the possibility of annihilating into gluons \cite{Isgur:1976qg}\cite{DeRujula:1975qlm}.
Witten \cite{Witten:1979vv} and Veneziano \cite{Veneziano:1979ec} pointed out that the problem should be studied by the $1/N_c$ expansion, and in this limit they found a relation between the $\eta^\prime$ mass and the pure-gauge topological susceptibility:
\begin{equation}\label{eq:chiWV}
\chi_{PG} = \frac{f_\pi^2 m_{\eta^\prime}^2}{2 n_f}
\end{equation}
where $f_\pi$ is the pion decay constant (normalized such that $f_\pi\sim 92$ MeV) and $n_f$ is the number of flavors. 
Including a finite quark mass, the relation becomes $\chi = \frac{f_\pi^2}{2 n_f} (m_{\eta^\prime}^2+m_\eta^2-2 m_{K}^2)$ \cite{Veneziano:1979ec}. Witten's argument is that the singlet and octet states are degenerate in the large $N_c$ limit, while a mass difference is produced by quark-antiquark annihilation diagrams at $\mathcal{O}(1/N_c)$. 

The $\eta^\prime$ mass has been computed with chiral effective Lagrangians by including an effective term that contains the topological charge density \cite{He:2009sb}\cite{Nekrasov:1996kt}\cite{Christos:1984tu}. 
In \cite{Novikov:1979ux} a purely gluonic current has been studied with QCD sum rules and the mass and the residue of the lowest-lying resonance have been computed, claiming that in the $\eta^\prime$ wave function both quark and gluon components are present. 
Other QCD sum rules studies include \cite{Singh:2018yvt}\cite{Forkel:2003mk}.
The $\eta^\prime$ mass and decay constant have been computed in lattice QCD, see, \emph{e.g.}, \cite{Ottnad:2017bjt} and references therein.

In the last years there were studies dealing with the $\eta^\prime$ problem using the gauge/gravity duality. 
In the AdS/CFT correspondence a four-dimensional gauge theory can be mapped to a gravity theory in the $AdS_5\times S_5$ space, such that gauge-invariant operators are described by proper higher-dimensional fields. 
Conformal symmetry requires that the mass of a field is fixed by the conformal dimension of the dual operator, and so it forces a vector field dual to a dimension-3 current to be a massless field. Hence, a global symmetry in the four-dimensional theory becomes a gauge symmetry on the gravity side.
However, since the QCD axial singlet current is not conserved, we expect that its dual field, \emph{i.e.} the singlet axial-vector field, gets a mass. 
In \cite{Klebanov:2002gr} the anomalous $U(1)$ symmetry becomes a spontaneously broken symmetry in supergravity, and the gauge field acquires a mass through the Higgs mechanism. 
In particular, a mass term for the gauge field dual to the $U(1)_R$ current is generated by a 2-form in the gravity theory which is not invariant under $U(1)_R$. 
On these bases, studies of the $U(1)_A$ problem in top-down \cite{Bigazzi:2019eks}\cite{Bartolini:2016dbk}\cite{Leutgeb:2019lqu}\cite{Brunner:2016ygk}\cite{Barbon:2004dq}\cite{Babington:2003vm} and bottom-up models \cite{Katz:2007tf}\cite{Schafer:2007qy}\cite{Hong:2009zw}\cite{Arean:2016hcs} have been proposed.
In \cite{Gursoy:2014ela} an axion field has been introduced in order to renormalize the anomalous conductivities, playing the role of the 2-form.
In \cite{Jimenez-Alba:2014iia} anomaly related transport phenomena in the presence of a magnetic field are studied with a massive vector field dual to a nonconserved current.
The mass term for the dimension-three vector field has been included by a gauge-invariant St\"uckelberg action. 
In \cite{Arean:2016hcs} a bottom-up model is used to compute the topological susceptibility and the mass of the singlet pseudoscalar meson in the Veneziano large-$N_c$ limit. 
In this model the QCD $\theta$ parameter is the source of a RR bulk field, and the $\eta^\prime$ mass emerges by the St\"uckelberg mechanism \cite{Casero:2007ae}.
The model has been used in \cite{Hamada:2020phg} for studying glueball spectra.

This issue has not been approached in the soft-wall model \cite{Karch:2006pv}.
Following  \cite{Arean:2016hcs} and \cite{Gursoy:2014ela}, we investigate the pseudoscalar singlet sector in the soft-wall model, and compute the topological susceptibility in pure gauge and for finite quark masses, and the $\eta^\prime$ mass.
A field dual to the $G\tilde G$ operator is introduced, and its mixing with the fields dual to the $\bar q \gamma_\mu\gamma_5 q$ and $\bar q \gamma_5 q$ currents is obtained by requiring $U(1)_A$ invariance of the Lagrangian for specific transformation rules.
In this scheme, the $\eta^\prime$ results from coupled equations of motion containing the three $0^{-+}$ isosinglet fields, such that its mass is increased with respect to the $\eta$ mass by a gluonic contribution. 
We also show that the anomaly equation naturally arises.
No mixing between $\eta$ and $\eta^\prime$ has been considered.
The dependence of the topological susceptibility on the quark mass is compared to lattice data and with the phenomenological prediction proposed in \cite{DiVecchia:1980yfw}\cite{Leutwyler:1992yt}, and the known asymptotic behaviours are reproduced. 

The paper is organized as follows. In Section \ref{sec:model} the soft-wall model for chiral-symmetry breaking is reviewed, and a scalar field $Y(x,z)$ containing the field dual to the QCD $G\tilde G$ operator is introduced, contributing to the description of the singlet pseudoscalar sector. 
In Section~\ref{sec:PG} the model is studied in the pure-gauge limit, in which only pseudoscalar glueballs are present, and the parameters introduced through the $Y(x,z)$ field are fixed by the coefficient of the glueball two-point function at high momentum and by the topological susceptibility. 
In Section~\ref{sec:eta8} nonsinglet pseudoscalar mesons are studied, and all the remaining parameters are fixed by the meson mass and decay constant. 
In particular, the mass of the up quark and the chiral condensate are fixed from the pion decay constant, and by matching the mass of the pseudoscalar ground state to the pion mass. 
Separately, the strange quark mass is fixed by matching the mass of the pseudoscalar ground state to the $\eta$ mass, while using the same value of $\langle \bar q q\rangle$ as before.
In Section~\ref{sec:eta0} the holographic version of the anomaly equation is found, reproducing the relation among the involved one-point functions in QCD; the $\eta^\prime$ mass is also obtained, in agreement with the experimental one. 
In the last Section the topological susceptibility is computed for different values of the quark mass, and the Witten-Veneziano relation is found at order $1/N_c$ in the large $N_c$ limit. 
Numerical results for the topological susceptibility are compared to lattice data.

\section{Model}\label{sec:model}
We consider the Lagrangian introduced in  \cite{Karch:2006pv} to study chiral symmetry breaking in a bottom-up AdS/QCD model. 
The $5d$ AdS space is characterized by the metric:
\begin{equation}
ds^2=\frac{R^2}{z^2} (dt^2-d\bar x^2-d z^2)\,,
\end{equation}
with $R$ the AdS curvature radius and $z \geqslant 0$ the additional bulk coordinate; $z=0$ corresponds to the AdS boundary. 
We use Greek letters for Minkowski ($4d$) indices, and capital letters for AdS ($5d$) indices.
According to the AdS/CFT correspondence, the global $U(n_f)_L\times U(n_f)_R$ symmetry of QCD is promoted to a gauge invariance of the Lagrangian of the $5d$ gravity theory under $U(n_f)_L\times U(n_f)_R$. 
The $5d$ fields dual to the left and right currents $\bar q_{L/R} \gamma^\mu T^A q_{L/R}$ are the $p$-forms $A_L(x,z)$ and $A_R(x,z)$ having $p=1$ and a zero $5d$ mass, as it results using the relation $m_5^2 R^2=(\Delta-p)(\Delta+p-4)$, where $\Delta$ is the conformal dimension of the $4d$ operator. 
Such massless fields, dual to conserved currents, are the gauge fields of the model. 
$T^A$, with $A=0,...,n_f^2-1$, are the generators of $U(n_f)$, with $T^0=\frac{1}{2n_f} I_{n_f}$ and $\mbox{Tr}(T^a T^b)=\delta^{ab}/2$ for $a,b=1,...,n_f^2-1$. The left and right gauge fields are expressed as $A^M_L(x,z)=A^{M,A}_L(x,z) T^A$. 
Vector and axial-vector fields are defined by $V=(A_L+A_R)/2$ and $A=(A_L-A_R)/2$, respectively. 
The $\bar q_R q_L$ operator is dual to the $5d$ tachyon field $X(x,z)=e^{i\eta^A(x,z) T^A} X_0(z) e^{i\eta^A(x,z) T^A}$, $\eta^A$ describing the nine pseudoscalar mesons. 
We work with $n_f=3$ quarks with equal masses $m_q$, so the expectation value of $X(x,z)$ is proportional to the identity matrix: $X_0(z)=\sqrt{2}v_q(z)\, I_{n_f}$.
The covariant derivative acts on $X(x,z)$ as
\begin{gather}
D_M X = \partial_M X +i [X,V_M]-i\{X,A_M\}\,.
\end{gather}
We choose the gauge $A_5=0$ and split the four components $A_\mu$ of the axial field in a transverse ($A_\perp$) and a longitudinal part ($\varphi$): $A_\mu^A = A_{\perp\, \mu}^A + \partial_\mu \varphi^A$.

To study $U(1)_A$, we also include a field $a(x,z)$ dual to the QCD operator $Q=\frac{\alpha_s}{8\pi} G_{\mu\nu}\tilde G^{\mu\nu}$. $Q$ appears in the QCD Lagrangian through the term $\theta \frac{g^2}{16 \pi^2} \mbox{Tr}(G_{\mu\nu}\tilde G^{\mu\nu})$, with $\theta$ a parameter. Therefore, we introduce the complex scalar field $Y(x,z) = Y_0(z)e^{2ia(x,z)}$, dual to the square of the gluon field strength, where the phase $a(x,z)$ is dual to the operator $Q$, sourced by $\theta$, and $Y_0(z)$ represents the vacuum expectation value of $Y(x,z)$, similarly to $X_0$ for the $X$ field. 
In top-down approaches the gluon contribution to the $U(1)_R$ anomaly has been discussed in \cite{Klebanov:2002gr}, showing that the anomaly arises because the Ramond-Ramond (RR) 2-form $C_2$ is not invariant under the $U(1)$ symmetry.
Such idea has been reproduced in a bottom-up approach \cite{Casero:2007ae}, where the coupling of RR form fields with the tachyon $X$ and the axial gauge field $A_M$ has been obtained from the Wess-Zumino action. 
The result involving the axion $a$, \emph{i.e.} the field sourced by $\theta$ that couples to $G\wedge G$ on the boundary, is gauge invariant if the following transformations of the 5$d$ fields under $U(1)_A$ are assumed:
\begin{gather}
\eta^0 \to \eta^0 -\alpha \\
\varphi^0 \to \varphi^0 -\alpha \\
a \to a - V_a \alpha \,. \label{eq:aU1}
\end{gather}
where $\alpha$ is the gauge parameter, and $V_a(z)$ is a potential term depending on the tachyon vacuum expectation value $v_q(z)$, such that $V_a\to 1$ as $z\to 0$ and $V_a\to 0$ for $v_q(z)\to\infty$. 
This guarantees that the transformation of the $a(x,z)$ field becomes a shift of the $\theta$ parameter on the boundary, which is the same result of an axial rotation in QCD.
We also assume that the field $a(x,z)$ appearing in the scalar field $Y(x,z)$ transforms as in \eqref{eq:aU1} under $U(1)_A$. 
This can be understood in a phenomenological construction noticing that $a(x,z)$ is a field comprising a term invariant under $U(1)_A$ ($a_{P}(x,t)$) plus a contribution proportional to the potential $V_a(z)$ that vanishes in pure gauge: $a(x,z)=a_{P}(x,z)+V_a(z) a_f(x,z)$, with $a_f$ transforming under $U(1)_A$ as $a_f \to a_f -\alpha$. 
In the model without $V_a$ the topological susceptibility is not affected by the quark mass, although it is still possible to get the mass of the $\eta^\prime$ by a proper set of parameters. \\
Assuming the transformation rule \eqref{eq:aU1}, a kinetic term $|\partial (Y_0 e^{2 i a})|^2$ would spoil gauge invariance of the Lagrangian.
We then write the kinetic term in a gauge-invariant way as in the Wess-Zumino action \cite{Casero:2007ae}:
\begin{equation}\label{eq:kineticY}
\mathcal{K}_a=|\partial_M Y_0(z) + 2 i Y_0 (\partial_M a(x,z) - \eta^0(x,z) \partial_M V_a(z) - A_M^0(x,z) V_a(z))|^2 \,.
\end{equation}
The full Lagrangian then reads:
\begin{equation}\label{eq:toymodel}
\mathcal{L}=\frac{1}{k}\sqrt{g}\, e^{-\phi}\,\left[ \mbox{Tr}\left\{ -\frac{1}{4g_5^2} (F_L^2+F_R^2) + |DX|^2 - m_X^2 |X|^2\right\} +\frac{1}{2} \mathcal{K}_a \right]\,.
\end{equation}
$\phi=c^2 z^2$ is the background dilaton field, with $c$ a mass parameter, introduced in the soft-wall model to break conformal invariance. 
The quadratic dependence of the dilaton has been chosen in order to get linear Regge trajectories in meson spectra \cite{Karch:2006pv}. 
From the relation between the $5d$ field mass and the conformal dimension $\Delta$ of the operator, the mass of the field $X(x,z)$ is $m_X^2=-3 R^2$.
The model with Lagrangian given by Eqs.~\eqref{eq:kineticY} and \eqref{eq:toymodel} describes two simultaneous mechanisms giving mass to the axial singlet field: a Higgs mechanism in the flavour sector and a St\"uckelberg mechanism in the gluon sector. \\
With these definitions, the Lagrangian \eqref{eq:toymodel} for pseudoscalar fields, up to quadratic order in fields $\varphi$, $\eta$, $a$, becomes
\begin{eqnarray}\label{eq:final-Lagrangian}
\mathcal{L} &=& \frac{R}{k} e^{-\phi}\left[ \frac{1}{4 n_f g_5^2 z}\, (\partial_z \partial_\nu \varphi^0)^2 + \frac{1}{2g_5^2 z}\, (\partial_z \partial^\nu \varphi^8)^2\right.\nonumber\\
&& -\frac{2 R^2\, v_q^2}{n_f z^3} (\partial_z \eta^0)^2+\frac{2 R^2\, v_q^2}{n_f z^3} (\partial_\nu \eta^0-\partial_\nu \varphi^0)^2-\frac{4\, R^2\, v_q^2}{z^3} (\partial_z \eta^8)^2+\frac{4\, R^2\, v_q^2}{z^3} (\partial_\nu \eta^8-\partial_\nu \varphi^8)^2\nonumber\\
&& \left.-\frac{2 R^2}{z^3} Y_0^2 (\partial_z a-\eta^0 \partial_z V_a)^2+\frac{2 R^2}{z^3}Y_0^2 (\partial_\nu a-V_a \partial_\nu \varphi^0)^2\right]\,.
\end{eqnarray}
$R/k=N_c/16 \pi^2$ and $g_5^2=3/4$ are fixed from scalar meson and vector meson two-point functions \cite{Colangelo:2008us}. $\varphi^8$ and $\varphi^0$ are the longitudinal component of the $A^8$ and $A^0$ axial-vector fields, respectively; $\eta^8$ describes the $\eta_8$ mesons, belonging to the octet representation of $SU(3)$, $\eta^0$ describes the singlet states. 
We are not considering the mixing between $\eta_8$ and $\eta_0$, so we identify $\eta_8$ with the $\eta$ and $\eta_0$ with the $\eta^\prime$.
We use
\begin{equation}\label{eq:vqsol}
 v_q(z)=\frac{m_q}{R} z+\frac{\sigma}{R} z^3
\end{equation}
as in the hard-wall model \cite{Erlich:2005qh}. 
This solution for $v_q(z)$ cannot be obtained from the equation of motion coming from the Lagrangian \eqref{eq:toymodel}, and a potential term must be added to get such a result.
We assume \eqref{eq:vqsol} without seeking which potential can give such a dependence in the soft-wall model. 
As an example, a different version of the soft-wall model, introducing a potential for $v_q(z)$, has been proposed in \cite{Ghoroku:2005vt}.
As we shall see, $v_q(z)$ in \eqref{eq:vqsol} is able to describe an explicit breaking of $SU(3)_A$ symmetry by the quark mass $m_q$, which produces a finite mass for pseudo-Goldstone bosons, and a spontaneous symmetry breaking through the quantity $\sigma$, proportional to the quark condensate. 
Such a relation can be obtained by deriving the on-shell action with respect to $m_q$, getting $\langle \bar q q\rangle= -\frac{N_c}{2 \pi^2} \sigma $.
As in  \cite{Arean:2016hcs} we use
\begin{equation}
V_a(z)=e^{-v_q(z)^2}\,,
\end{equation}
having the requested behaviour in the $z\to 0$ limit.
We also use 
\begin{equation}\label{eq:Y0general}
Y_0(z)=\frac{y_0}{R}+\frac{2\, y_1}{R\, c^4} (e^{c^2 z^2} (-1 + c^2 z^2)+1)\,,
\end{equation}
with asymptotic behaviour
\begin{equation}\label{eq:Y0SWzto0}
Y_0(z)\xrightarrow[z\to 0]{} \frac{1}{R} \left(y_0 + y_1 z^4 + ...\right) \,.
\end{equation}
This is the solution to the equation of motion coming from the Lagrangian term $\mathcal{L}_y=-\frac{R^3 e^{-c^2z^2}}{2\, k\, z^3}  Y_0'(z)^2$, which is included in \eqref{eq:toymodel}.

\section{Pure gauge}\label{sec:PG}
The starting point to understand pseudoscalar glueball and meson mixing in the soft wall is the theory with no quarks. 
In this case the Lagrangian reads 
\begin{eqnarray}\label{eq:PSglueballLag}
\mathcal{S}_{PG} &=& \frac{R}{k}\int d^5 x \, e^{-\phi}\, \left[ -\frac{2 R^2}{z^3} Y_0^2 (\partial_z a_P)^2+\frac{2 R^2}{z^3}Y_0^2 (\partial_\mu a_P)^2\right]\,.
\end{eqnarray}
Pseudoscalar glueballs have been also studied in AdS/QCD frameworks by a different choice of the $5d$ field \cite{Hashimoto:1998if}\cite{Csaki:1998qr}. 
The bulk-to-boundary propagator, solution to the equation of motion obtained from Eq.~\eqref{eq:PSglueballLag} with boundary condition $\tilde a_P(0)=1$ and giving a finite action ($\tilde a_P\xrightarrow[z\to\infty]{}0$), is
\begin{equation}\label{eq:BTBPaPG}
\tilde a_P(z)=\frac{y_0}{Y_0(z)} \Gamma\left(2-\frac{q^2}{4c^2}\right)  U\left(-\frac{q^2}{4c^2},-1,c^2 z^2\right) \,,
\end{equation}
where $U$ is the Tricomi confluent hypergeometric function, and $\Gamma$ is the Riemann Gamma function.
The eigenvalues are $m_n^2=4c^2(n+2)$, therefore in this model pseudoscalar glueballs are degenerate with scalar glueballs \cite{Colangelo:2007pt}. Using $c=388$ MeV, as obtained by fitting the $\rho$ meson mass \cite{Karch:2006pv}, the lightest state has mass $m_{G\tilde G,0}=1.1$ GeV. 

The two-point correlation function is obtained in the holographic framework by  deriving twice the on-shell action with respect to the sources of the operator under consideration. For pseudoscalar glueballs it is given by
\begin{equation}\label{eq:2pfPSglue}
\Pi_{aa}(q^2) = \left. \frac{R}{k} \frac{4 Y_0(z)^2 e^{-c^2 z^2}}{z^3} \tilde a'_P(z) \tilde a_P(z)\right|_{z\to 0} \,.
\end{equation}
The poles of the two-point function are located at $q^2=m_n^2$, and the residues are $R_n=\frac{R}{k} 32 c^6 y_0^2\, (n+1)\, (n+2)$. From the relation $R_n=m_n^4\, f_n^2$ \cite{Gabadadze:1997zc}, we get the decay constants $f_n^2=\frac{R}{k} 2c^2 y_0^2\, \frac{(n+1)}{(n+2)}$. 

The high-$Q^2$ expansion ($Q^2=-q^2$) of the two-point function of the pseudoscalar glueball operator in QCD is \cite{Gabadadze:1997zc}
\begin{equation}
\Pi_{QCD}(Q^2)=-\left( \frac{\alpha_s}{8\pi}\right)^2  Q^4 \left( \frac{2}{\pi^2} \log Q^2 + {\cal O}(Q^{-4}) \right)\,.
\end{equation}
Likewise, the two-point function in Eq.~\eqref{eq:2pfPSglue} in the limit $Q^2=-q^2\to\infty$ is:
\begin{equation}
\Pi_{aa}(Q^2)= Q^4 \left( -\frac{N_c}{32 \pi^2} y_0^2 \log Q^2 + {\cal O}(1) \right)\,.
\end{equation}
Matching the two expressions at leading order we find 
\begin{equation}
y_0=\frac{1}{\sqrt{N_c}}\frac{\alpha_s}{\pi}\,.
\end{equation}
The last relation shows that the $Y$ field appears in the full Lagrangian at a lower order ($\mathcal{O}(1/N_c)$) in the large $N_c$ expansion with respect to the other fields, as expected \cite{Witten:1998uka}.
It is then convenient to emphasize the $1/N_c$ factor redefining
\begin{equation}
\hat Y_0(z)=\sqrt{N_c} Y_0(z)= \frac{\hat y_0}{R}+\frac{2\, y_1}{R\, c^4} (e^{c^2 z^2} (-1 + c^2 z^2)+1)
\end{equation} 
where $\hat y_0=\frac{\alpha_s}{\pi}$.
Using $\hat y_0=\frac{1}{\pi}$, we find for the ground state $f_{G\tilde G,0}=9.8$ MeV.

Considering the candidates for these states, the pseudoscalar $\eta(1405)$ has some features that can indicate a glueball component \cite{Close:1996yc}, \emph{i.e.} it has not been produced in $\gamma\gamma$, it has large branching ratios in $J/\psi$ decays, and has not been seen to decay to $\gamma\gamma$. 
Some models support the hypothesis of $\eta(1405)$ being a glueball, while lattice QCD and QCD sum rules predict masses larger than 2 GeV \cite{Forkel:2003mk}\cite{Gabadadze:1997zc}\cite{Chen:2005mg}\cite{Morningstar:1999rf}\cite{Mathieu:2008me}. 
In \cite{Faddeev:2003aw} an argument supporting a degeneracy between scalar and pseudoscalar glueballs has been put forward.
In the present model, if the mass scale $c$ appearing in the dilaton is fixed from the $\rho$ mass, low masses for pseudoscalar glueballs are found, and the nearest state to $\eta(1405)$ is the first radial excitation, with $m_{G\tilde G,1}=1.34$ GeV.

The topological susceptibility is the second derivative of the vacuum energy with respect to $\theta$ at $\theta=0$.
It is different from zero both in pure gauge and in the theory with physical quarks, while it vanishes if there is at least a massless quark, so in this case the theory has no $\theta$ dependence.
In the AdS/QCD model its definition becomes \cite{Katz:2007tf}
\begin{equation}\label{eq:defchi}
\chi_t=-\lim_{q^2\to 0} \Pi_{aa}(q^2)\,.
\end{equation}
Using Eqs.~\eqref{eq:BTBPaPG} and \eqref{eq:2pfPSglue} for $q^2=0$, in pure gauge we find
\begin{equation}\label{eq:finalchiPG}
\chi_{PG}=\frac{1}{\pi^2} \hat y_0 y_1  = \frac{1}{\pi^2} \frac{\alpha_s}{\pi} y_1\,.
\end{equation}
From the Witten-Veneziano relation in \eqref{eq:chiWV}, we obtain $\chi_{PG}\sim (191 \mbox{ MeV})^4$, a value confirmed by independent analyses \cite{Vicari:2008jw}\cite{DelDebbio:2004ns}\cite{Teper:1999wp}\cite{Alles:1996nm}. 
In this model it corresponds to $\hat y_0=1/\pi$ and $y_1=0.041 \mbox{ GeV}^4$. 
For the slope of the topological susceptibility at zero momentum, connected to the spin content of the proton \cite{Narison:1998aq}, we find
\begin{equation}
\chi^\prime_{PG}=\frac{1}{8\pi^2} c^2 \hat y_0^2 (1-\gamma_E+\log{\nu^2/c^2})  \,,
\end{equation}
where $\nu$ is the renormalization scale; for $\nu=1$ GeV, corresponding to subtracting only the $\log{\nu^2}$ term, we have $\sqrt{\chi^\prime_{PG}}=0.021$ GeV.

\section{Nonsinglet pseudoscalar mesons}\label{sec:eta8}
Let us now consider the pseudoscalar meson octet.
The equations of motion for $\varphi^8$ and $\eta^8$ can be obtained from the Lagrangian \eqref{eq:final-Lagrangian}:
\begin{equation}\label{eq:eom1eta8}
\partial_z \left( \frac{v_q^2  e^{-\phi}}{z^3} \partial_z \eta^8\right)+ q^2 \frac{v_q^2 e^{-\phi}}{z^3} (\eta^8-\varphi^8)=0 
\end{equation}
\begin{equation}\label{eq:eom2eta8}
\partial_z \left( \frac{e^{-\phi}}{g_5^2 z} \partial_z \varphi^8\right)+ \frac{8v_q^2 e^{-\phi}}{z^3} (\eta^8-\varphi^8)=0 \,.
\end{equation}
Combining the two equations and integrating, the following relation can be obtained:
\begin{equation}\label{eq:eom3eta8}
\frac{q^2}{g_5^2 z} \partial_z \varphi^8- \frac{8 v_q^2}{z^3} \partial_z \eta^8 =0 \,,
\end{equation}
where the integration constant is zero since $\partial_z\varphi^8$ and $\partial_z\eta^8$ vanish as $z\to\infty$ due to boundary conditions.
Following the matrix formalism of \cite{Kaminski:2009dh}, we aggregate the two fields in the vector 
\begin{equation}
  \Phi = \left(
  \begin{array}{c}
  \varphi^8\\
  \eta^8
  \end{array}
  \right)\,,
  \end{equation}
so the Lagrangian  \eqref{eq:final-Lagrangian} containing nonsinglet fields reads:
\begin{equation}
\mathcal{L} = \frac{R}{k}  e^{-\phi} \left( {\Phi'}^\dagger B \Phi' + {\Phi}^\dagger C \Phi \right)
\end{equation}
with
\begin{equation}
B = \left(
\begin{array}{cc}
\frac{q^2}{2 g_5^2 z} & 0\\
0 & -\frac{4 v_q^2 R^2}{z^3}
\end{array}
\right)\,,
\end{equation}
\begin{equation}
C = \frac{4 q^2 v_q^2 R^2}{z^3} \left(
\begin{array}{cc}
1 & -1\\
-1 & 1
\end{array}
\right)\,.
\end{equation}
Operator mixing on the boundary implies that the fields in the bulk are given by a linear combination of the sources in Fourier space \cite{Kaminski:2009dh}:
\begin{equation}
\Phi = F\, \Phi_0 \,,
\end{equation}
where $\Phi_0(q^2) = (\varphi_0^8(q^2),-\eta_0^8(q^2))^T$\footnote{The minus sign in the $\eta^8$ source comes from the condition $\eta^8\to-\eta^8_0$ on the boundary \cite{Erlich:2005qh}} is the vector containing the sources of the two operators, so  $F(z,q^2)$ tends to the identity matrix on the boundary $z\to \varepsilon$. Moreover, we require $\partial_z F\to 0$ for $z\to \infty$ in order to guarantee finiteness of the action.
Standing the arbitrariness of the sources $\Phi_0$, the equations of motion for $F$ are
\begin{equation}
 \partial_z \left( e^{-\phi} B\, F' \right)- e^{-\phi} C \, F =0 \,,
\end{equation}
or, in components,
\begin{eqnarray}
&& \partial_z \left( e^{-\phi} B_{00} F'_{00} \right)-e^{-\phi} (C_{00} F_{00}+C_{01} F_{10})=0\\
&& \partial_z \left( e^{-\phi} B_{00} F'_{01} \right)-e^{-\phi} (C_{00} F_{01}+C_{01} F_{11})=0\\
&& \partial_z \left( e^{-\phi} B_{11} F'_{10} \right)-e^{-\phi} (C_{10} F_{00}+C_{11} F_{10})=0\\
&& \partial_z \left( e^{-\phi} B_{11} F'_{11} \right)-e^{-\phi} (C_{10} F_{01}+C_{11} F_{11})=0 \,.
\end{eqnarray}
The on-shell action reads:
\begin{equation}
\mathcal{S}_{os} = -\lim_{z\to \varepsilon} \frac{R}{k}  \int d^4 k \,\, e^{-c^2z^2} \, \Phi_0^\dagger {F}^\dagger B F' \Phi_0 \,,
\end{equation}
and the one-point functions are defined by
\begin{equation}
 \renewcommand{\arraystretch}{1.2}
  J^{8} = \left. \frac{\partial \mathcal{S}_{os}}{ \partial \Phi_0}\right|_{z\to \varepsilon} = \left(
  \begin{array}{c}
  \langle J^{8}_{\varphi}\rangle \\
  \langle J^{8}_{\eta}\rangle
  \end{array}
  \right) \,,
  \end{equation}
where $\langle J^{8}_{\varphi}\rangle$ and $\langle J^{8}_{\eta}\rangle$ are the one-point functions of the longitudinal axial current ($\partial_\mu \bar\psi\gamma_5\gamma^\mu T^8 \psi$) and the pseudoscalar current ($ 2m_q\bar\psi\gamma_5 T^8\psi$), respectively, in the nonsinglet sector.
In the soft-wall model, after imposing boundary conditions, we find
\begin{eqnarray}
  \langle J^{8}_{\varphi}\rangle &=& -\left. \frac{R}{k} \frac{e^{-\phi} q^2 }{2 g_5^2 z} (\varphi^8)' \right|_{z\to \varepsilon}\\
  \langle J^{8}_{\eta}\rangle &=& -\left. \frac{R}{k} \frac{e^{-\phi} 4 v_q^2 R^2 }{z^3} (\eta^8)' \right|_{z\to \varepsilon} \,,
  \end{eqnarray}
so, using Eq.~\eqref{eq:eom3eta8}, we find 
\begin{equation}
\langle J^{8}_{\varphi}\rangle=\langle J^{8}_{\eta}\rangle \,,
\end{equation}
which establishes the partial conservation of axial current.

The two-point functions are defined by
\begin{equation}
 \renewcommand{\arraystretch}{1.5}
\Pi^{88} = \left. \frac{\partial^2 \mathcal{S}_{os}}{\partial \Phi_0 \partial \Phi_0}\right|_{z\to \varepsilon} = \left(
\begin{array}{cc}
\Pi^{88}_{\varphi\varphi} & \Pi^{88}_{\varphi\eta}\\
\Pi^{88}_{\eta\varphi} & \Pi^{88}_{\eta\eta}
\end{array}
\right) \,,
\end{equation}
and are given by
\begin{eqnarray}
\Pi^{88}_{\varphi\varphi}  &=& -\lim_{z\to \varepsilon} \frac{R}{k} \frac{q^2 e^{-\phi}}{g_5^2 z} F_{00}' F_{00}  \\
\Pi^{88}_{\varphi\eta} = \Pi^{88}_{\eta\varphi} &=&  \lim_{z\to \varepsilon} \frac{R}{k} e^{-\phi}\left(\frac{q^2}{2 g_5^2 z} F_{01}' F_{00} - \frac{4 v_q^2 R^2}{z^3} F_{10}' F_{11} \right)  \\
\Pi^{88}_{\eta\eta}  &=&  \lim_{z\to \varepsilon} \frac{R}{k} e^{-\phi}  \frac{8 v_q^2 R^2 }{z^3} F_{11}' F_{11} 
\end{eqnarray}
where we have already required that off-diagonal matrix elements of $F$ vanish on the boundary.
We find $\Pi^{88}_{\varphi\varphi} = \Pi^{88}_{\varphi\eta}=\Pi^{88}_{\eta\varphi}= \Pi^{88}_{\eta\eta}$.

In QCD, the first term in the large $Q^2=-q^2$ expansion of the two-point function of the longitudinal component of the axial-vector current is \cite{narisonbook}\footnote{An additional $1/2$ factor has been included to take into account the definition of the axial field used here, with $\mbox{Tr}(T^aT^b)=\delta^{ab}/2$.}
\begin{equation}
\psi_5(Q^2) \xrightarrow[Q^2\to\infty]{} \frac{3}{16\pi^2} 4 m_q^2 \left(1+\frac{11}{3}\frac{\alpha_s}{\pi}\right) Q^2\log Q^2+\mathcal{O}(Q^{-2})\,.
\end{equation}
In the soft-wall model the leading order result agrees with the QCD one: 
\begin{equation}
\Pi_{\varphi\varphi}^{88}(Q^2) \xrightarrow[Q^2\to\infty]{} \frac{3}{16\pi^2} 4 m_q^2 Q^2\log Q^2+\mathcal{O}(Q^{2})\,.
\end{equation}

In all the numerical computations,  $z\geqslant\varepsilon=0.001$ GeV$^{-1}$ has been used.
Since $X_0$ is proportional to the identity matrix, the equations of motion of the fields belonging to the octet are all equal, and the corresponding mesons are degenerate.
Fixing $|\langle q\bar q\rangle| = (0.281~\mathrm{GeV})^3$, by requiring that the first eigenvalue of Eqs.~\eqref{eq:eom1eta8}-\eqref{eq:eom2eta8} is $m_{\eta_8} = m_\pi = 139~\mathrm{MeV}$, we get $m_q = m_u = 3.7~\mathrm{MeV}$. 
Requiring $m_{\eta_{8}} = m_\eta = 548~\mathrm{MeV}$ instead, we get $m_q = m_s = 59.5~\mathrm{MeV}$.
The same meson masses are also found from the first pole of the two-point functions.
Using the spectral representation
\begin{equation}\label{eq:spectralrep}
\Pi_{\varphi\varphi}^{88}(q^2)=\sum_n \frac{m_n^4 f_n^2}{m_n^2-q^2}+P_2(q^2)\,,
\end{equation}
where $m_n$ and $f_n$ are the masses and decay constants of the nonsinglet pseudoscalar mesons, $P_2(x)$ is a polynomial of degree two, we find $f_{0,u}=92.3$ MeV and $f_{0,s}=103$ MeV, with $f_{0,s}/f_{0,u}=1.12$. For $m_q=0$, the pion-decay constant is
\begin{equation}\label{eq:pionDC}
f_\pi^2=\left.- \frac{R}{k}\frac{e^{-\phi}}{g_5^2 z} \partial_z F_{00} \right|_{\begin{subarray}{l}
q^2=0\\
z\to \varepsilon
\end{subarray}}=(91.6 \mbox{ MeV})^2\,.
\end{equation}
Our numerical data agree with the  Gell-Mann-Oakes-Renner mass formula, which reads
$
m_\pi^2 = 2 m_u |\langle \bar q q \rangle| / f_\pi^2 + \mathcal{O}(m_u^2)
$, 
where $f_\pi$ is the pion decay constant in the chiral limit. 
The Gell-Mann-Oakes-Renner relation has been recovered in the hard-wall model in \cite{Erlich:2005qh}, and a similar derivation also holds in the soft-wall model.
The same result for the decay constant in the chiral limit is found from the following relation:
\begin{equation}\label{eq:pionDCwave}
f_\pi^2=\left.- \frac{R}{k}\frac{e^{-\phi}}{g_5^2 z} \partial_z \psi \right|_{z\to \varepsilon}=(91.6 \mbox{ MeV})^2\,,
\end{equation}
which has been introduced in \cite{Grigoryan:2007wn} in the hard-wall model, where $\psi=\varphi^8-\eta^8$ is the solution to Eq.~\eqref{eq:eom2eta8} at $q^2=0$ with $\eta^8(z)=-1$ and with boundary conditions $\psi(z)=1$ at $z=\varepsilon$ and $\psi'(z)=0$ for $z\to\infty$.

\section{Singlet pseudoscalar mesons}\label{sec:eta0}
If $a(x,z)=0$, then $\eta^0(x,z)$ and $\eta^8(x,z)$ satisfy the same equation of motion with the same boundary conditions, and singlet and nonsinglet states have the same mass.
If $a(x,z)\neq 0$, then $\eta^0(x,z)$ and $a(x,z)$ are coupled, while the equations for $\eta^8(x,z)$ do not change. 

The equations of motion for $a,\varphi^0,\eta^0$ are
\begin{equation}\label{eq:neweom-eta0}
\partial_z \left( \frac{v_q^2  e^{-\phi}}{n_f z^3} \partial_z \eta^0\right)+ q^2 \frac{v_q^2 e^{-\phi}}{n_f z^3} (\eta^0-\varphi^0) + \frac{\hat Y_0^2 e^{-\phi}}{N_c z^3} (\partial_z V_a) (\partial_z a -\eta^0 \partial_z V_a)=0 
\end{equation}
\begin{equation}\label{eq:neweom-phi0}
\partial_z \left( \frac{e^{-\phi}}{2 n_f g_5^2 z} \partial_z \varphi^0\right)+\frac{4v_q^2 R^2 e^{-\phi}}{n_f z^3} (\eta^0-\varphi^0) + \frac{4 \hat Y_0^2 R^2 e^{-\phi}}{N_c z^3} V_a (a- \varphi^0 V_a)=0 \,
\end{equation}
\begin{equation}\label{eq:neweom-a}
\partial_z \left( \frac{\hat Y_0^2 e^{-\phi}}{z^3} (\partial_z a -\eta^0 \, \partial_z V_a)\right)+q^2 \frac{\hat Y_0^2 e^{-\phi}}{z^3} (a- \varphi^0 V_a)=0 \,.
\end{equation}
A combination of the three equations gives (after integrating)
\begin{equation}\label{eq:neweom-tot}
 \frac{4 v_q^2 R^2 }{n_f z^3} \partial_z \eta^0 -  \frac{q^2 }{2 n_f g_5^2 z} \partial_z \varphi^0 +  \frac{4 \hat Y_0^2 R^2}{N_c z^3} V_a (\partial_z a -\eta^0 \partial_z V_a) =0 \,,
\end{equation}
where the integration constant is zero since $\partial_z\varphi^0$, $\partial_z\eta^0$, $a$, and $\hat Y_0^2V_a$ vanish as $z\to\infty$ due to boundary conditions.
The last equation can be also obtained by requiring the gauge-fixing condition $A_5(q,z)=0$.

In the matrix formalism \cite{Kaminski:2009dh} adopted in the previous section, the three fields are contained in the vector
\begin{equation}
  \Psi = \left(
  \begin{array}{c}
  \varphi^0\\
  \eta^0\\
  a
  \end{array}
  \right)\,,
  \end{equation}
and the Lagrangian  \eqref{eq:final-Lagrangian} containing singlet fields reads
\begin{equation}
  \mathcal{L} = \frac{R}{k} e^{-\phi} \left( {\Psi'}^\dagger M \Psi' + {\Psi'}^\dagger N_1 \Psi+ {\Psi}^\dagger N_2 \Psi' + {\Psi}^\dagger P \Psi \right)
\end{equation}
with
\begin{equation}
M = \left(
\begin{array}{ccc}
\frac{q^2}{4 n_f g_5^2 z} & 0 & 0\\
0 & \frac{-2 v_q^2 R^2}{n_f z^3} & 0\\
0 & 0  & \frac{-2 \hat Y_0^2 R^2}{N_c z^3}
\end{array}
\right)\,,
\end{equation}
\begin{equation}
N_1 = \frac{2 \hat Y_0^2 R^2 V'_a}{N_c z^3}\left(
\begin{array}{ccc}
0 & 0 & 0\\
0 & 0 &  0\\
0 & 1  & 0
\end{array}
\right)\,,
\end{equation}
\begin{equation}
N_2 = \frac{2 \hat Y_0^2 R^2 V'_a}{N_c z^3} \left(
\begin{array}{ccc}
0 & 0 & 0\\
0 & 0 &  1\\
0 & 0  & 0
\end{array}
\right)\,,
\end{equation}
\begin{equation}
\renewcommand{\arraystretch}{2.5}
 P = R^2 \left(
\begin{array}{ccc}
\frac{2 q^2 v_q^2}{n_f z^3}+ \frac{2 \hat Y_0^2 q^2 V_a^2}{N_c z^3} & \frac{-2 q^2 v_q^2}{n_f z^3} & \frac{-2 \hat Y_0^2 q^2 V_a}{N_c z^3}\\
\frac{-2 q^2 v_q^2}{n_f z^3} & \frac{2 q^2 v_q^2}{n_f z^3}- \frac{2 \hat Y_0^2(V_a')^2}{N_c z^3} & 0 \\
 \frac{-2 \hat Y_0^2 q^2 V_a}{N_c z^3} & 0 & \frac{2 \hat Y_0^2 q^2}{N_cz^3}
\end{array}
\right)\,.
\end{equation}
The equations of motion are
\begin{equation}
  \partial_z \left( e^{-\phi} (M \Psi'+N_1 \Psi \right)- e^{-\phi}(N_2 \Psi'+P \Psi) =0 \,.
\end{equation}
Also here, because of operator mixing, the fields in the bulk are given by a linear combination of the sources in Fourier space:
\begin{equation}
  \Psi = H\, \Psi_0 \,,
\end{equation}
where $\Psi_0(q^2) = (\varphi_0^0(q^2),-\eta_0^0(q^2),a_0(q^2))^T$ is the vector containing the sources of the three operators, and $H(z,q^2)$ tends to the identity matrix on the boundary $z\to \varepsilon$. 
Moreover, finiteness of the action implies that, as $z\to\infty$, $\partial_z H\to 0$ for the first two rows, and  $H\to 0$ for the third row.
For the arbitrariness of the sources $\Psi_0$, the equations of motion for $H$ are
\begin{equation}\label{eq:eomH}
  \partial_z \left( e^{-\phi} (M\, H'+N_1 H) \right)- e^{-\phi}(N_2\, H'+P \, H) =0 \,.
\end{equation}
The on-shell action reads
\begin{equation}
\mathcal{S}_{os} = - \lim_{z\to \varepsilon} \frac{R}{k}\int d^4 k \,\, e^{-\phi} \, (\Psi_0^\dagger {H}^\dagger M H' \Psi_0 +\Psi_0^\dagger {H}^\dagger N_1 H \Psi_0)\,,
\end{equation}
and the two-point functions are
\begin{equation}\label{eq:tpf00}
 \renewcommand{\arraystretch}{1.5}
 \Pi^{00}  = \left. \frac{\partial^2 \mathcal{S}_{os}}{\partial \Psi_0 \partial \Psi_0}\right|_{z\to \varepsilon} = \left(
\begin{array}{ccc}
\Pi^{00}_{\varphi\varphi} & \Pi^{00}_{\varphi\eta}  & \Pi_{\varphi a}\\
\Pi^{00}_{\eta\varphi} & \Pi^{00}_{\eta\eta} & \Pi_{\eta a}\\
\Pi_{a\varphi} & \Pi_{a\eta} & \Pi_{a a}\\
\end{array}
\right) \,.
\end{equation}

From the on-shell action we compute the one-point functions:
\begin{equation}
  \renewcommand{\arraystretch}{1.5}
  J^{0} = \left. \frac{\partial \mathcal{S}_{os}}{ \partial \Psi_0}\right|_{z\to \varepsilon} = \left(
  \begin{array}{c}
  \langle J^{0}_{\varphi}\rangle \\
  \langle J^{0}_{\eta}\rangle\\
  \langle J^{0}_{a}\rangle
  \end{array}
  \right) \,,
  \end{equation}
where $\langle J^{0}_{\varphi}\rangle$ and $\langle J^{0}_{\eta}\rangle$ are the one-point functions of the longitudinal axial current ($\partial_\mu \bar\psi\gamma_5\gamma^\mu T^0 \psi$) and the pseudoscalar current ($ 2m_q\bar\psi\gamma_5 T^0\psi$), respectively, in the singlet sector, while $ \langle J^{0}_{a}\rangle$ is the one-point function of the topological charge density ($\langle\frac{\alpha_s}{8\pi}G_{\mu\nu}\tilde G^{\mu\nu}\rangle$).
In the soft-wall model, after imposing boundary conditions, we find
\begin{eqnarray}
  \langle J^{0}_{\varphi}\rangle &=&  -\left. \frac{R}{k} \frac{e^{-\phi} q^2}{4 n_f g_5^2 z} \varphi'_0\right|_{z\to \varepsilon} \\
  \langle J^{0}_{\eta}\rangle &=& -\left. \frac{R}{k} \frac{e^{-\phi} 2 v_q^2 R^2}{n_f z^3}\eta'_0\right|_{z\to \varepsilon} \\
  \langle J^{0}_{a}\rangle &=& \left. \frac{R}{k} \frac{e^{-\phi} 2 \hat Y_0^2 R^2}{N_c z^3}a' \right|_{z\to \varepsilon} \,.
  \end{eqnarray}
These are related by a Ward identity obtained taking the $z\to \varepsilon$ limit of Eq.~\eqref{eq:neweom-tot}:
\begin{equation}\label{eq:anomaly1pt}
-\langle J^{0}_{\eta}\rangle + \langle J^{0}_{\varphi}\rangle + \langle J^{0}_{a}\rangle =0 \,,
\end{equation}
which is the holographic representation of the QCD anomaly equation:
\begin{equation}
  \partial_\mu J_A^\mu =  2m_q \bar\psi\gamma_5 T^0\psi - \frac{\alpha_s}{4\pi}\,G_{\mu\nu}\tilde G^{\mu\nu}\,\mathrm{tr}T^0 \,\,.
\end{equation}
Notice that in this model the singlet axial and pseudoscalar currents have been defined using the $U(1)_A$ generator $T^0=\frac{1}{2 n_f} I_{n_f}$.

\begin{figure}[h]
\begin{center}
\includegraphics[width=8cm]{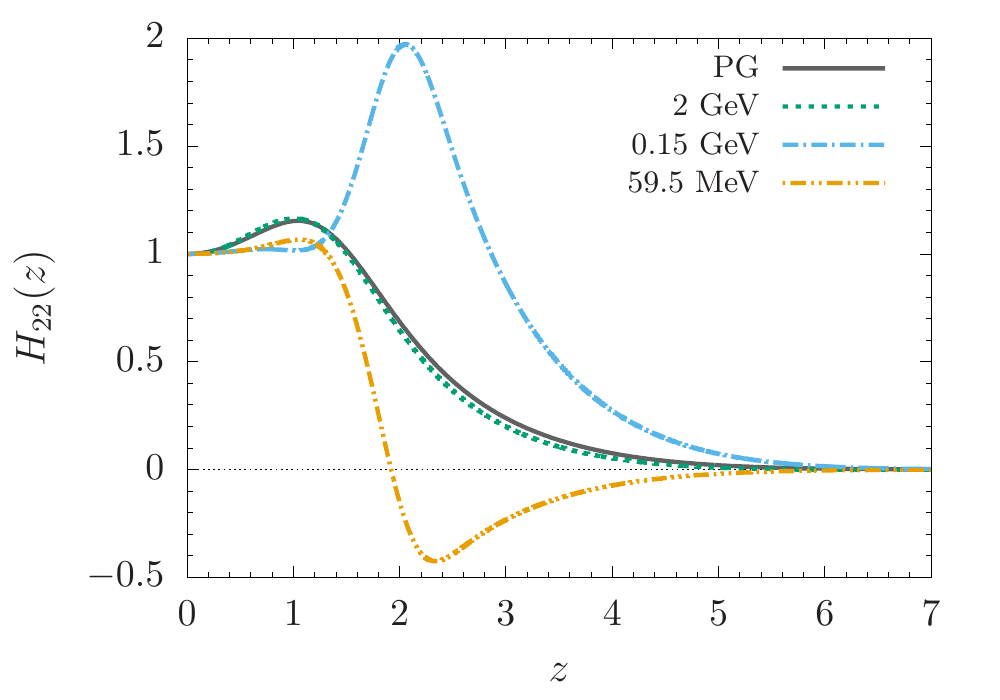}
\caption{Bulk-to-boundary propagator of the field $a_P(z)$ in pure gauge (PG) at $q^2=1$ GeV$^2$ in Eq.~\eqref{eq:BTBPaPG} (solid grey line); $H_{22}(z)$ at $q^2=1$ GeV$^2$ and for the indicated values of the quark mass. At large quark masses $H_{22}(z)$ approaches the pure-gauge solution.}
\label{fig:BTBP}
\end{center}
\end{figure}

In Fig.~\ref{fig:BTBP}, we show the effect of the meson-glueball mixing on the element $H_{22}$, which is the only nonvanishing element of the matrix $H$ in pure gauge. 
It is also compared with the bulk-to-boundary propagator of pseudoscalar glueballs in Eq.~\eqref{eq:BTBPaPG}.

\begin{figure}[h]
\begin{center}
\includegraphics[width=8cm]{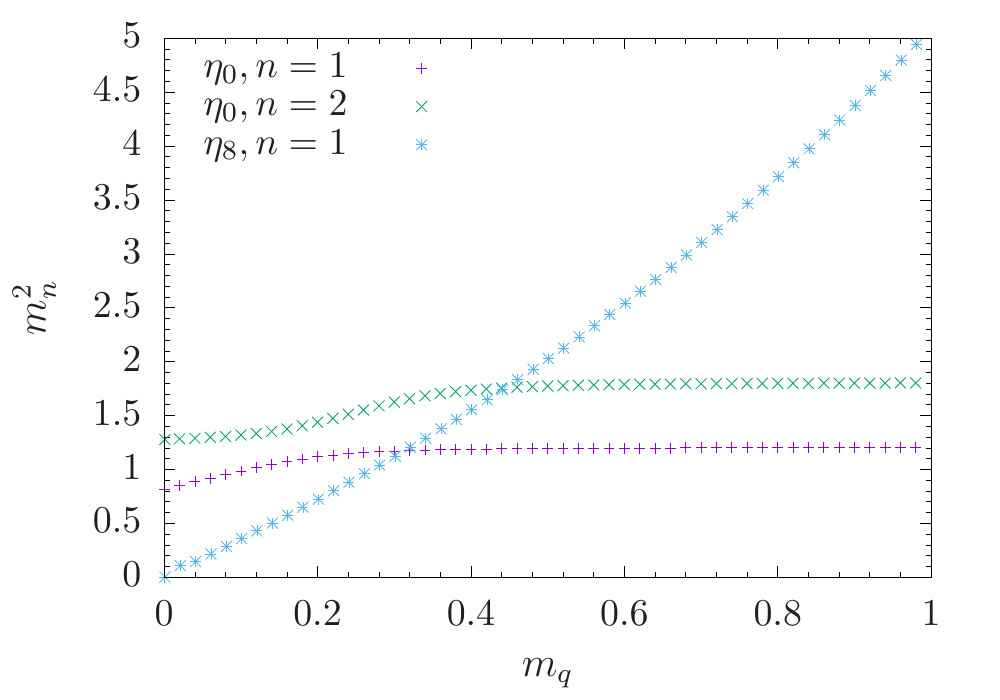}
\caption{Squared mass (GeV$^2$) of the octet ground state, of the ground state and first radial excitation of the singlet field versus the quark mass (GeV). }
\label{fig:mass}
\end{center}
\end{figure}

With the parameters used in Sections \ref{sec:PG} and \ref{sec:eta8}, we find that the mass of the first pole of the two-point correlation functions is $m_{\eta^\prime}= 958$ MeV for $m_q=m_s=59.5$ MeV, while in the chiral limit it is $m_{\eta^\prime}=903$ MeV. In the $1-2.3$ GeV mass range there are many resonances, generated by the meson-glueball mixing, with the following masses: 1.14, 1.37, 1.57, 1.75, 1.91, 2.05, 2.15, and 2.24 GeV.
According to the way the singlet state is constructed in this model, the mass of the first pole approaches the mass of the lightest glueball for high quark masses, as shown in Fig.~\ref{fig:mass}.

Seven $0^{-+}$ states with isospin 0 are listed in the Particle Data Group compilation, \emph{i.e.} $\eta$, $\eta^\prime$, $\eta(1295)$, $\eta(1405)$, $\eta(1475)$, $\eta(1760)$ (not established), $\eta(2225)$ (not established), and one state with unknown isospin, \emph{i.e.} $X(1835)$ (not established). According to this model, all higher-mass states can be radial excitations of $\eta^\prime$, comprising quark and gluon components. 
However, a thorough analysis of pseudoscalar-meson spectroscopy should require the inclusion of a mass difference between strange and up/down quarks, and the $\eta_0$-$\eta_8$ mixing.

\section{Topological susceptibility}
The topological susceptibility $\chi_t$ in the full theory with quarks is
\begin{equation}\label{eq:chit-full}
\chi_{t} = -\Pi_{aa}(q^2=0) = -\left. \frac{R}{k} \frac{4 \hat Y_0^2 R^2 e^{-\phi}}{N_c z^3} H'_{22} H_{22} \right|_{\substack{q^2\to 0\\ z\to 0}} \,,
\end{equation}
with $\Pi_{aa}(q^2)$ defined in Eq.~\eqref{eq:tpf00}, and $H_{22}$ solution of Eq.~\eqref{eq:eomH}. 
In particular, $H_{22}$ is only coupled to the matrix elements $H_{02}$ and $H_{12}$ by the following equations:
\begin{eqnarray}
\label{eq:H02}\partial_z\left( \frac{e^{-\phi}}{2 g_5^2 n_f z} H^\prime_{02}\right) - e^{-\phi} \frac{4 v_q^2 R^2}{n_f z^3} (H_{02}-H_{12}) + e^{-\phi} \frac{4 \hat Y_0^2 R^2 V_a}{N_c z^3} (H_{22}-H_{02} V_a) =0\\
\label{eq:H12}\partial_z\left( \frac{e^{-\phi} v_q^2}{n_f z^3} H^\prime_{12}\right) - e^{-\phi} \frac{q^2 v_q^2}{n_f z^3} (H_{02}-H_{12}) + e^{-\phi} \frac{\hat Y_0^2 V'_a}{N_c z^3} (H'_{22}-H_{12} V'_a) =0\\
\label{eq:H22}\partial_z\left( \frac{e^{-\phi} \hat Y_0^2}{z^3} (H^\prime_{22}-H_{12} V'_a)\right) + e^{-\phi} \frac{q^2 \hat Y_0^2}{z^3} (H_{22}-H_{02} V_a)  =0 \,.
\end{eqnarray}

In the chiral limit, the Witten-Veneziano relation can be obtained both in the $N_c\to\infty$ and in the $n_f/N_c\to 0$ limits \cite{Giusti:2001xh}. The parameters involved in this model scale as $R/k\sim \mathcal{O}(N_c)$,  $m_{\eta^\prime}\sim \mathcal{O}(1/\sqrt{N_c})$ or $m_{\eta^\prime}\sim \mathcal{O}(\sqrt{n_f/N_c})$, while the others are $\mathcal{O}(1)$.
By expanding Eqs.~\eqref{eq:neweom-eta0}-\eqref{eq:neweom-tot} in $1/N_c$ for $q^2=m_{\eta^\prime}^2$ one finds at lowest order:
\begin{eqnarray}
\label{eq:phi0-zero} \partial_z \left( \frac{e^{-\phi}}{2 n_f g_5^2 z} \partial_z (\varphi^0)_0 \right)-\frac{4v_q^2 R^2 e^{-\phi}}{n_f z^3} ((\varphi^0)_0-(\eta^0)_0)=0 \,
\\
\partial_z \left( \frac{v_q^2  e^{-\phi}}{n_f z^3} \partial_z (\eta^0)_0\right)=0 \\
\label{eq:a-zero} \partial_z \left( \frac{\hat Y_0^2 e^{-\phi}}{z^3} ( \partial_z(a)_0-(\eta^0)_0 V'_a) \right)=0 \\
\label{eq:tot0-zero}\frac{m_{\eta^\prime}^2 }{2 n_f g_5^2 z} \partial_z  (\varphi^0)_0 =  \frac{4 \hat  Y_0^2 R^2}{N_c z^3} V_a  (\partial_z (a)_0-(\eta^0)_0 V'_a)  \,,
\end{eqnarray}
where the subscript $0$ in $(\varphi^0)_0$, $(\eta^0)_0$  and $(a)_0$ means that they are the zeroth-order ($\mathcal{O}(1)$) solutions to the equations of motion. 
The second equation tells us that $(\eta^0)_0$ is constant, in particular in the chiral limit $(\eta^0)_0=-1$ \cite{Grigoryan:2007wn}.
The first and third equations are decoupled. 
In particular, Eq.~\eqref{eq:phi0-zero} coincides with the equation of motion of the nonsinglet field $\psi$ introduced at the end of Section \ref{sec:eta8} in the chiral limit at $q^2=0$, which is related to the pion decay constant by Eq.~\eqref{eq:pionDCwave}.
Equation~\eqref{eq:a-zero}, after defining $\tilde a_P(z,q^2)=a(z,q^2)+V_a(z)$, coincides with the equation of motion of the pseudoscalar glueball in pure gauge at $q^2=0$.
%
Then, after multiplying both sides of Eq.~\eqref{eq:tot0-zero} by $(R/k) e^{-\phi}$ and taking the limit $z\to 0$, from Eqs.~\eqref{eq:pionDCwave} and \eqref{eq:chit-full}, it results in 
\begin{equation}
 \frac{m_{\eta^\prime}^2 f_\pi^2}{2 n_f} =  \chi_{PG}  \,,
\end{equation}
which is the Witten-Veneziano relation.

\begin{figure}[h]
\begin{center}
\includegraphics[width=8cm]{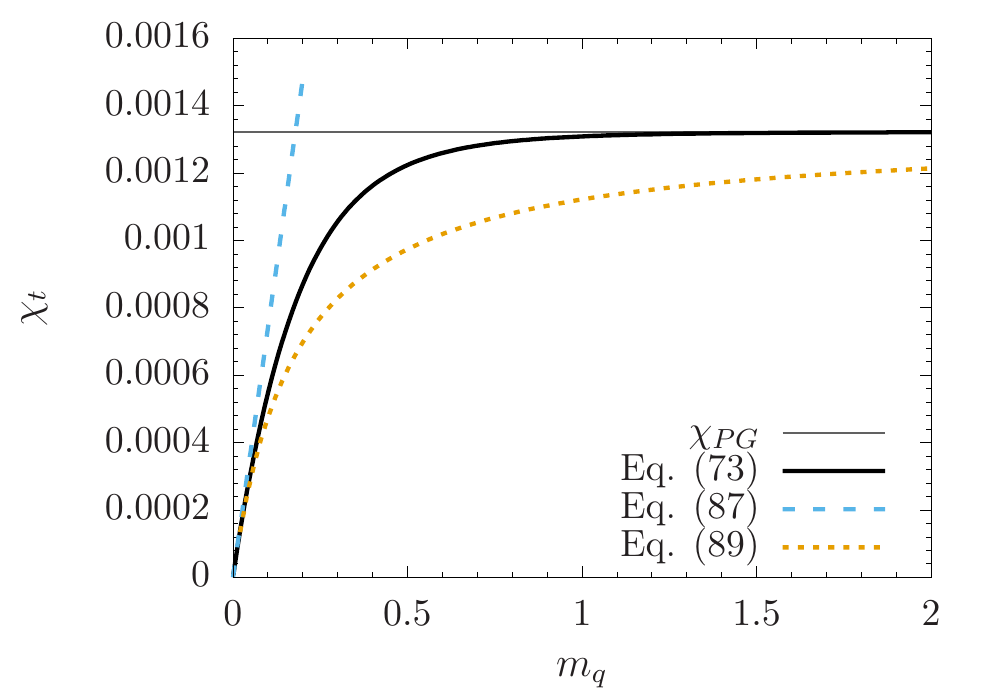}
\caption{Topological susceptibility (GeV$^4$) at different values of the quark mass (GeV) computed in this holographic model from Eq.~\eqref{eq:chit-full} or \eqref{eq:chit-sum}-\eqref{eq:chif} (black curve); topological susceptibility in pure gauge given by Eq.~\eqref{eq:finalchiPG} (horizontal grey line); expected topological susceptibility at low quark mass in Eq.~\eqref{eq:chilowmq} (dashed cyan line); predicted topological susceptibility from Eq.~\eqref{eq:chileut} (dotted orange line).}
\label{fig:chi}
\end{center}
\end{figure}

Let us compute the topological susceptibility as a function of the quark mass, and express it in a more useful form. 
Eqs.~\eqref{eq:H12} and \eqref{eq:H22} at $q^2=0$ can be integrated obtaining
\begin{eqnarray}
e^{-\phi} \frac{4 \hat Y_0^2 R^2}{N_c z^3} (H'_{22}-V'_a H_{12}) = A_1 \,,\\
e^{-\phi} \frac{4 v_q^2 R^2}{n_f z^3} H'_{12} + V_a A_1 =0 \,,
\end{eqnarray}
where the integration constant is nonzero in the former case and zero in the latter, according to the behaviour of the functions at $z\to \infty$.
One can then write
\begin{eqnarray}
-1 &=& \int_0^\infty dz \,  H_{22}'(z) = A_1 \int_0^\infty dz \, \frac{ e^{\phi(z)} z^3 }{4} \left(\frac{N_c }{\hat Y_0(z)^2 R^2} + n_f \left(\frac{V_a(z)}{ v_q(z) R}\right)^2 \right)\,.
\end{eqnarray}
From $\chi_t=-\frac{R}{k}\, A_1$, one obtains
\begin{equation}\label{eq:chit-sum}
\frac{1}{\chi_t}  =\frac{1}{\chi_{PG}}+\frac{1}{\chi_f}
\end{equation}
with 
\begin{equation}\label{eq:chif}
\frac{1}{\chi_f}=\frac{k}{R} n_f  \int_0^\infty dz \, \frac{ e^{\phi(z)} z^3 }{4 }  \left(\frac{ V_a(z)}{ v_q(z) R}\right)^2 \,.
\end{equation}
The values of the topological susceptibility versus the quark mass are shown in Fig.~\ref{fig:chi}. At small quark mass we find
\begin{equation}\label{eq:chilowmq}
  \chi_t \xrightarrow[m_q\to 0]{} \chi_f \sim  \frac{\langle \bar q q\rangle}{n_f} m_q\,, 
\end{equation}
as predicted in chiral perturbation theory \cite{Vicari:2008jw}.
The behaviour in Eq.~\eqref{eq:chilowmq} is represented by the dashed line in Fig~\ref{fig:chi}.
At large quark mass we find $\chi_f \xrightarrow[m_q\to\infty]{} \frac{3}{\pi^2 n_f} m_q^4$, so the pure-gauge value (horizontal line in Fig.~\ref{fig:chi}) is recovered in the limit $m_q \to \infty$ :
\begin{equation}
\chi_t \xrightarrow[m_q\to \infty]{} \chi_{PG} \,, 
\end{equation}
a result that has been numerically checked for very large $m_q$.    
The result in Fig.~\ref{fig:chi} is similar to the plot on the left of Fig. 7 of \cite{Arean:2016hcs}. 
This is an indication that the profile of the topological susceptibility is mostly related to the Wess-Zumino holographic description of the $U(1)_A$ problem represented by Eq.~\eqref{eq:kineticY}, while it is less affected by the other features of the underlying model, as, in particular, how conformal and chiral symmetries are broken, or the use of a dynamical background.

\begin{figure}[h]
\begin{center}
\includegraphics[width=8cm]{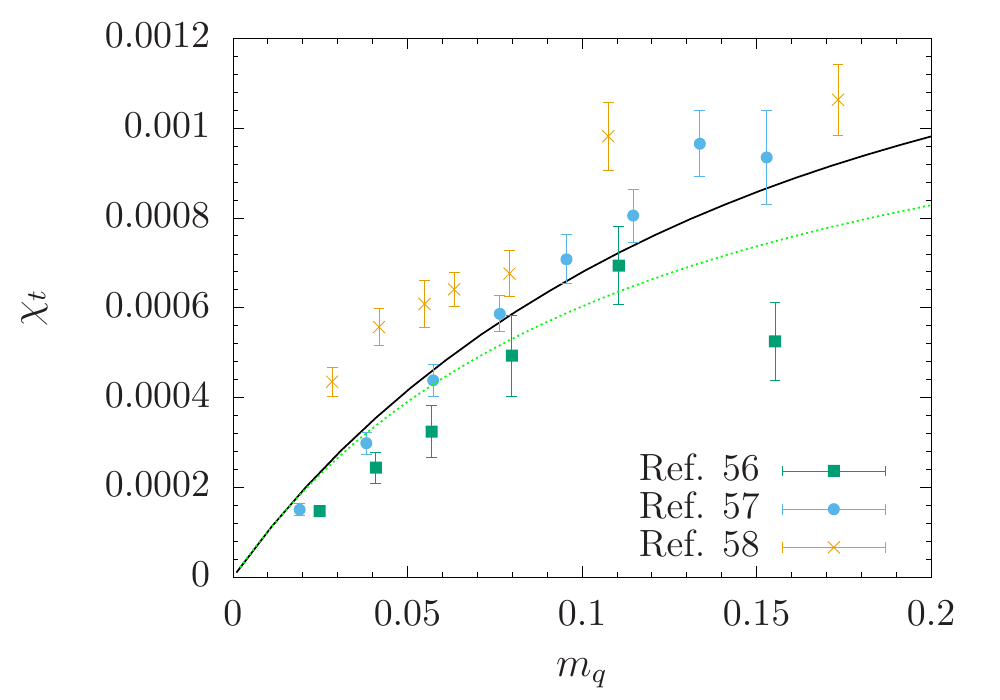}
\caption{Lattice data from Refs. \cite{Aoki:2007pw}\cite{Chiu:2011dz}\cite{DeGrand:2020utq}, and the topological susceptibility computed from Eq.~\eqref{eq:chit-full} (solid black line) and from Eq.~\eqref{eq:chileut}  (dashed green line), both curves at $n_f=2$ and $|\langle \bar q q\rangle|=(0.281~\mathrm{GeV})^3$. Units are GeV on the horizontal axis, and GeV$^4$ on the vertical one.}
\label{fig:latticechi}
\end{center}
\end{figure}

The formula
\begin{equation}\label{eq:chileut}
\frac{1}{\chi_t}=\frac{1}{\chi_{PG}}+\frac{n_f}{m_q |\langle \bar q q\rangle|}
\end{equation}
interpolating between the two known asymptotic  behaviours was obtained in \cite{DiVecchia:1980yfw}\cite{Leutwyler:1992yt}.
The result in Eq.~\eqref{eq:chit-sum} shows that the topological susceptibility computed in this model gets the same expression as in Eq.~\eqref{eq:chileut}, and they numerically agree for low values of the quark mass, as shown in Fig.~\ref{fig:chi}, since a more complicated form of the second term in the right-hand side is needed to fit $\chi_f$ at higher $m_q$.
In Fig.~\ref{fig:latticechi} some lattice data with $n_f=2$ \cite{Aoki:2007pw}\cite{Chiu:2011dz}\cite{DeGrand:2020utq} are compared to $\chi_t$ computed from Eq.~\eqref{eq:chit-full}  and from Eq.~\eqref{eq:chileut} with $n_f=2$ and $|\langle \bar q q\rangle|=(0.281~\mathrm{GeV})^3$. 
Lattice data confirm the suppression of the topological susceptibility at decreasing quark mass.
It is difficult to carry out a quantitative comparison of lattice data with Eq.~\eqref{eq:chileut} or with the values found in this holographic model, since uncertainties involved in lattice simulations \cite{Vicari:2008jw} and the different values of the chiral condensate should be taken into account.

\section{Conclusions}
We have computed the masses of singlet and nonsinglet pseudoscalar mesons in the soft-wall holographic model of QCD. 
We have found that the mixing of singlet states with pseudoscalar glueballs can explain the large mass of the $\eta^\prime$. We have not considered the mixing between singlet and nonsinglet states with isospin zero.
The holographic version of the anomaly equation has been found in Eq.~\eqref{eq:anomaly1pt}. 
In this respect, it is worth emphasising that in this model partial conservation of axial current, the anomaly equation and the Witten-Veneziano relation are derived from the constraint equations \eqref{eq:eom3eta8} and \eqref{eq:neweom-tot}, obtained by a combination of the equations of motion of the involved fields.

A key result of this paper is the computation of the topological susceptibility $\chi_t$ in the soft-wall model for any value of quark mass in Eqs.~\eqref{eq:chit-sum}-\eqref{eq:chif}, and its comparison to lattice simulations at $n_f=2$. Moreover, our result agrees with the formula \eqref{eq:chileut} proposed in Refs. \cite{DiVecchia:1980yfw}\cite{Leutwyler:1992yt}, since we also find that the contributions from quarks and gluons are combined as a sum of their reciprocals. We have found that in the full topological susceptibility the correction $\chi_f$ to the pure-gauge value depends linearly on the quark mass for low values of $m_q$, as in chiral perturbation theory, while other corrections arise for higher $m_q$, in particular $\chi_f$ diverges as $m_q^4$ at infinite $m_q$.

These results have also interesting prospectives concerning the computation of the spectral functions of the $\eta^\prime$ at finite temperature, which have recently been object of lattice studies \cite{Kotov:2019dby}. In this respect, in \cite{Miranda:2009uw}\cite{Colangelo:2009ra}\cite{Bellantuono:2014lra} it has been shown that the study of spectral functions in the soft-wall model is numerically feasible and produces consistent results.

The main ingredient of this model is the potential $V_a(z)$, characterizing the transformation rule of the glueball field in the bulk under $U(1)_A$. 
We have fixed it from the Wess-Zumino action, but different choices  could be explored. 
The computation of the topological susceptibility shows that $V_a(z)$ has to vanish at infinity at least as $e^{-\phi}$ (where $\phi$ is the dilaton).

A related issue, not discussed here, is the strong \emph{CP} problem \cite{DiVecchia:2013swa}. 
The topological term in the QCD Lagrangian, proportional to the $\theta$ parameter, generating the gluonic contribution to the $\eta^\prime$ mass is not invariant under \emph{CP}. 
This term produces a nonzero neutron electric dipole moment, which is expected to be tiny. The experimental limit is $\theta < 10^{-10}$. 
Then, a fine-tuning problem arises, trying to understand why $\theta$ is so small, while naturally one would expect such parameter of $\mathcal{O}(1)$.
A solution to the strong \emph{CP} problem is the Peccei-Quinn mechanism. 
In \cite{Bigazzi:2019eks} and \cite{Cox:2019rro} it has been proposed how the Peccei-Quinn mechanism can be implemented in a top-down and bottom-up holographic model, respectively.

\section*{Acknowledgements}
We thank P.~Colangelo for helpful discussions and for pointing out unclear points and subtleties.

\end{document}